\documentclass[a4paper,12pt]{article}
\usepackage{standalone}

\usepackage[utf8]{inputenc}
\usepackage[english]{babel}
\usepackage{amssymb}
\usepackage{amsmath}
\usepackage{mathrsfs}
\usepackage{cite}
\usepackage{url}
\usepackage[ruled]{algorithm2e}
\usepackage{graphicx,xcolor}
\usepackage{gnuplottex}
\usepackage{float}

\makeatletter

\newcommand{\redc}{\underline{\mathrm{REDC}}}

\newcommand*{\mydiv}{%
  \nonscript\mskip-\medmuskip\mkern5mu%
  \mathbin{\operator@font div}\penalty900\mkern5mu%
  \nonscript\mskip-\medmuskip
}
\newcommand*{\mymod}{%
  \nonscript\mskip-\medmuskip\mkern5mu%
  \mathbin{\operator@font mod}\penalty900\mkern5mu%
  \nonscript\mskip-\medmuskip
}
\newsavebox{\@brx}
\newcommand{\myllangle}[1][]{\savebox{\@brx}{\(\m@th{#1\langle}\)}%
  \mathopen{\copy\@brx\kern-0.5\wd\@brx\usebox{\@brx}}}
\newcommand{\myrrangle}[1][]{\savebox{\@brx}{\(\m@th{#1\rangle}\)}%
  \mathclose{\copy\@brx\kern-0.5\wd\@brx\usebox{\@brx}}}
\makeatother

\title{Speeding up decimal multiplication}
\author{Viktor Krapivensky}
\date{%
    November 25, 2020
}

\begin{document}

\maketitle

\begin{abstract}
    Decimal multiplication is the task of multiplying two numbers in base $10^N.$
    Specifically, we focus on the number-theoretic transform (NTT) family of algorithms.
    Using only portable techniques, we achieve a 3x---5x speedup over the \textbf{mpdecimal}
    library.
    In this paper we describe our implementation and discuss further possible optimizations.
    We also present a simple cache-efficient algorithm for in-place $2n \times n$ or
    $n \times 2n$ matrix transposition, the need for which arises in the ``six-step algorithm''
    variation of the matrix Fourier algorithm, and which does not seem to be widely known.
    Another finding is that use of two prime moduli instead of three makes sense even considering
    the worst case of increasing the size of the input, and makes for simpler answer recovery.
\end{abstract}

\section{Introduction}

Fast multiplication of large decimal numbers is of interest in computer algebra systems, arbitrary precision calculators like \textbf{bc}, and generally software that needs to present the result of some calculations in decimal form.
Since conversion from binary to decimal and vice versa takes $\Theta(\mathscr{M}(n) \log n)$ time, where $\mathscr{M}(n)$ is the time needed to multiply two numbers of size $n,$
it makes sense to keep the base decimal if the calculations themselves take $o(\mathscr{M}(n) \log n)$ time. Otherwise, it might be better to pay the cost of conversions and perform the calculations in binary,
as the binary form, being native for computers, has lower hidden constant. Also, as far as NTT is concerned, the binary form allows for better granularity when deciding how many ``digits'' to pack into a single word.

It is a common theme in practical analysis of algorithms that ``fancy'' algorithms are slow when $n$ is small.
For example, virtually all partical implementations of the ``fancy'' sorting algorithms that are divide-and-conquer in nature --- quick sort and merge sort --- do in fact fall back to ``dumb'' quadratic sorts if the size of the input is less than some threshold.
In some cases, it is not even a matter of ``fancy'' versus ``dumb'' dichotomy, but rather there is a ``hierarchy of fanciness'':
a group of algorithms with different asymptotic behavior such that it makes sense to use each one of them only for some range of values of $n.$
To give an example, \textbf{libgmp} employs the following ``hierarchy of fanciness'' for multiplication~\cite{gmp_fh}:
first goes the ``basecase'' (quadratic) algorithm; then, the Karatsuba algorithm; then, variations of the Toom-Cook algorithm (Toom-3; then Toom-4; then Toom-6.5; then Toom-8.5); finally, the Schönhage--Strassen algorithm (which is Fourier transform-based).

Indeed, for smaller sizes of the multiplicands, it does makes sense to use quadratic, Karatsuba or Toom-Cook algorithms;
as $n$ gets large, however, asymptotic considerations start to outweigh.
For multiplication, this means that different variations of the fast Fourier transform start to be used.

\section{Notation}

The following notation is employed throughout this document:
$$\langle x_0, \ldots, x_{n-1} \rangle [i] = x_i.$$
In order words, square brackets mean zero-based tuple indexing.

By $\mathbb{N}$ we mean $\{0, 1, 2, \ldots\}.$

By $x \mymod y,$ where $x \in \mathbb{Z},$ $y \in \mathbb{N} \setminus \{0\},$
we mean $z \in \mathbb{N} \cap [0; y-1]$ such that $z \equiv x \, (\mymod \, y).$

By $x \mydiv y,$ where $x \in \mathbb{N},$ $y \in \mathbb{N} \setminus \{0\},$
we mean $\lfloor x / y \rfloor.$

A ring is a set $\mathcal{R}$ equipped with two binary operations, $+$ and $\cdot,$ such that
$\langle \mathcal{R}, + \rangle$ is an abelian group,
$\cdot$ is associative and there an identity element with respect to $\cdot$ in $\mathcal{R},$
and $\cdot$ is both left distributive and right distributive with respect to $+.$

\section{Overview of the discrete Fourier transform}

The plan of this section is as follows:
\begin{itemize}

    \item First, we define the notion of cyclic convolution for an arbitrary algebraic ring.

    \item Next, we show how to reduce multiplication of two integers $x, \, y$ in base $B,$ such that
          $x \in [0; B^n - 1], \, y \in [0; B^m - 1],$
          to cyclic convolution of size $n+m$ (or of any larger size) over ring $\mathbb{Z}$ with some
          special properties: inputs to such a convolutions are in
          $[0; B-1]$
          and outputs are in
          $[0; (B-1)^2 \min\{n, m\}].$

    \item Then, we introduce discrete Fourier transform (DFT) and inverse discrete Fourier transform (IDFT)
          in their general form --- over an algebraic field $\mathcal{F}.$
          We show the connection between cyclic convolution and DFT/IDFT.
          We will learn that, employing fast Fourier transform (FFT) for DFT and IDFT, it is possible to
          compute cyclic convolutions over $\mathcal{F}$ efficiently.

    \item Finally, we discuss how different Fourier transform-based algorithms for integer
          multiplication employ different strategies to reduce the computation of cyclic
          convolutions over $\mathbb{Z}$ to cyclic convolutions over different algebraic fields.
\end{itemize}

\subsection{Cyclic convolution}

We now define the cyclic convolution.
Consider an aribtrary algebraic ring $\mathcal{R}.$
The cyclic convolution $x \star y, \, (\_ \star \_) \colon \mathcal{R}^n \times \mathcal{R}^n \to \mathcal{R}^n$ is defined as follows:
\begin{equation} \label{eq:star}
(x \star y)[k] = \sum\limits_{i=0}^{n - 1} x[i] y[(k - i) \mymod n].
\end{equation}

\subsection{Multiplication via cyclic convolution over $\mathbb{Z}$}

Let us now fix a base $B \in \mathbb{N}, \, B > 1,$ and introduce the following notation:
$$\myllangle x_0, \ldots, x_{n-1} \myrrangle = \sum\limits_{i=0}^{n-1} x_i B^i,$$
where $x_i \in \mathbb{N}, \, x_i < B.$

Let us say we want to multiply
$x = \myllangle x_0, \ldots, x_{n-1} \myrrangle$
and
$y = \myllangle y_0, \ldots, y_{m-1} \myrrangle.$

This is to say, we want to find the product
$z = xy = \myllangle z_0, \ldots, z_{n+m-1} \myrrangle.$

We compute the following cyclic convolution (over the ring of $\mathbb{Z}$):
$$\langle c_0, \ldots, c_{n+m-1} \rangle =
\langle x_0, \ldots, x_{n-1}, \underbrace{0, \ldots, 0}_{\text{$m$ zeros}} \rangle
\star
\langle y_0, \ldots, y_{m-1}, \underbrace{0, \ldots, 0}_{\text{$n$ zeros}} \rangle.$$

Note that
\begin{equation} \label{eq:ci_bound}
c_i \le (B-1)^2 \min\{n, m\}
\end{equation}
as both $x_i$ and $y_i$ are less that $B$ and, because of our zero padding, the summation
in~\eqref{eq:star} can have at most $\min\{n,m\}$ non-zero summands.

The representation of $z$ in base $B$ can then be recovered by the following iterative process:
\begin{subequations} \label{eq:sigmai_and_zi}
\begin{align}
\sigma_0 &= 0; \\
z_i &= (\sigma_i + c_i) \mymod B; \\
\sigma_{i+1} &= (\sigma_i + c_i) \mydiv B.
\end{align}
\end{subequations}

It would be useful to obtain some upper bound on the value of $\sigma_i.$

Define $L = (B-1)^2 \min\{n,m\}$ and remember that in~\eqref{eq:ci_bound}, we established that
$c_i \le L.$
Define also the following sequence:
$$a_0 = 0, \, a_{i+1} = (a_i + L) \mydiv B.$$

Clearly $\sigma_i \le a_i$ irrespective of the concrete values of $\langle c_0, \ldots, c_{n+m-1} \rangle.$

Define another sequence by
$$a^\prime_0 = 0, \, a^\prime_{i+1} = (a^\prime_i + L) / B.$$

Now $a^\prime_i \ge a_i$ and
$$a^\prime_i =
\sum\limits_{j=0}^{i-1} \frac{L}{B^{i - j}} =
\sum\limits_{j=1}^{i} \frac{L}{B^j} =
L \sum\limits_{j=1}^{i} \frac{1}{B^j}.$$

We have
\begin{equation} \label{eq:sigmai_bound}
\sigma_i \le a_i \le a^\prime_i < \lim\limits_{i \to \infty} a'_i = \frac{L}{B - 1} = (B-1)\min\{n, m\}.
\end{equation}

We have thus reduced multiplication of $n$-digit $x$ and $m$-digit $y$ to a convolution of size
$n+m.$ But note that, for any $k \in \mathbb{N},$
$$\myllangle x_0, \ldots, x_{n-1} \myrrangle = \myllangle x_0, \ldots, x_{n-1}, \underbrace{0, \ldots, 0}_{\text{$k$ zeros}} \myrrangle,$$
so we can as well reduce it to a convolution of any size $n+k+m \ge n+m$ by padding $x$ (or, similarly, $y$)
with $k$ higher zeros and ignoring the higher $k$ elements of the resulting vector.

\subsection{Discrete Fourier transform}

We fix an algebraic field $\mathcal{F}$ and the length of transform, $N \in \mathbb{N}.$

We will pretend natural numbers are in $\mathcal{F}$ by adopting the following identity:
$$n = \underbrace{\widehat{1} + \widehat{1} + \cdots + \widehat{1}}_\text{$n$ times},$$
where $\widehat{1} \in \mathcal{F}$ is the multiplicative identity in $\mathcal{F};$ the degenerate case of this identity is $0 = \widehat{0},$
where $\widehat{0} \in \mathcal{F}$ is the additive identity in $\mathcal{F}.$

We require that there exists a primitive $N$-th root of unity in $\mathcal{F}$:
such $\xi \in \mathcal{F}$ that $\xi^N = 1$ and $\xi^k \ne 1$ for every $0 < k < N.$
We fix one such $\xi.$

We also require that $N$ be invertible (i.e., non-zero) in $\mathcal{F}.$

Discrete Fourier transform $\mathfrak{f} \colon \mathcal{F}^N \to \mathcal{F}^N$ is then defined as follows:
\begin{equation} \label{eq:dft}
\mathfrak{f}(x)[k] = \sum\limits_{i=0}^{N-1} x[i] \xi^{ik}.
\end{equation}

Inverse discrete Fourier transform $\mathfrak{f}^{-1} \colon \mathcal{F}^N \to \mathcal{F}^N$ is defined similarly,
up to the negated exponent of $\xi$ and multiplication by $N^{-1}$:
\begin{equation} \label{eq:idft}
\mathfrak{f}^{-1}(x)[k] = N^{-1} \sum\limits_{i=0}^{N-1} x[i] \xi^{-ik}.
\end{equation}

It is easy to see that both $\mathfrak{f}$ and $\mathfrak{f}^{-1}$ are linear, which means that for any
$\alpha \in \mathcal{F}$ and $x \in \mathcal{F}^N,$
\begin{subequations} \label{eq:transform_linearity}
\begin{align}
\mathfrak{f}(\alpha x) &= \alpha \mathfrak{f}(x); \\
\mathfrak{f}^{-1}(\alpha x) &= \alpha \mathfrak{f}^{-1}(x).
\end{align}
\end{subequations}
Above, the product of a scalar and a vector
$\alpha v = \alpha \langle v_0, \ldots, v_{n-1} \rangle$ denotes, as usual,
$\langle \alpha v_0, \ldots, \alpha v_{n-1} \rangle.$

The convolution theorem~\cite{conv_theorem} says that, for any two vectors $x, y \in \mathcal{F}^N,$
\begin{equation} \label{eq:conv_theorem}
x \star y = \mathfrak{f}^{-1}\big( \mathfrak{f}(x) \cdot \mathfrak{f}(y) \big),
\end{equation}
where $\cdot$ denotes scalar (element-wise) product.

Note that $\star$ is bilinear, which means that it is linear in both of its arguments:
for any scalar $\alpha \in \mathcal{F}$ and any two vectors $x, y \in \mathcal{F}^N,$
\begin{equation} \label{eq:star_linearity}
x \star (\alpha y) = (\alpha x) \star y = \alpha (x \star y).
\end{equation}

There are algorithms for computing both discrete Fourier transform and inverse discrete Fourier transform
efficiently --- in $O(N \log N)$ time, provided that addition and multiplication
in $\mathcal{F}$ take $O(1)$ time. Any such algorithm is called a fast Fourier transform (FFT).

Although an $O(N \log N)$ FFT algorithm exists that works for arbitrary $N,$ even
prime~\cite{bluestein},
the most widely used FFT algorithm, the Cooley---Tukey algorithm, works by re-writing a transform
of composite length $N = N_1 N_2$ into smaller transforms of lengths $N_1$ and $N_2.$
Thus, it works only for the values of $N$ which are highly composite;
particularly, if $N$ is a power of two, its variants named
decimation in time (DIT) and decimation in frequency (DIF), can be used.

\subsection{Complex FFT using floating point}

The obvious approach to calculate a cyclic convolution over $\mathbb{Z}$ is to pick a field that
contains $\mathbb{Z}$ and has primitive $n$-th roots of unity for any $n$ --- namely, $\mathbb{C},$ the
field of complex numbers.
A primitive $n$-th root of unity can then be expressed by the formula $e^{2 \pi i / n}.$
This is the core of the ``FFT multiplication'' algorithm usually taught in classes:
approximate $\mathbb{C}$ with a pair of double-precision floating point values, compute the
convolution and round the results back to integers.

The main drawback of this algorithm is precision issues.
Formally, it is not even a sound algorithm for multiplication: provided that the precision of
your floating-point values is bounded, you can not multiply arbitrarily large numbers with it: at
some point, you are going to get a round-off error large enough to produce a wrong digit in the
answer~\cite{precision_paper}.

In order to get a provably correct answer, you have to put less bits of information into the
floating-point values than otherwise possible
(tables of maximum number of decimal/binary digits are given in
\cite[p. 560--561]{mattcomp} and \cite{gmp_paper};
for explicit formulas, see \cite{precision_paper}).
Because of that, in order to achieve performance
parity with multiplication algorithms that abstain from use of floating point, platform-specific
SIMD extensions are often used~\cite{gmp_paper}.
Also, in order to get a provably correct result, you need to know the precision of your
floating-point types, as well as possible quirks of their implementation --- i.e., depend on the
platform.

\subsection{Number-theoretic transform}

For any prime number $p,$ there exists the finite field $\mathrm{F}_p$ of integers modulo
$p.$ The number-theoretic transform is defined as the discrete Fourier transform in
$\mathrm{F}_p,$ for some prime $p.$
Note that $\mathrm{F}_p$ contains a primitive $n$-th root of unity if and only if $n$ divides $(p-1).$

We will now discuss the application of number-theoretic transform to integer multiplication.

First note that, in practice, it is always possible to obtain a reasonable upper bound on the length
of numbers we would ever need to multiply.
Let us thus say we have chosen the maximum length of a multiplicand, $M \in \mathbb{N}.$

One approach is to pick a prime $p > (B-1)^2 M$ and perform computations in
$\mathrm{F}_p.$
Remember that the inputs to the convolution over $\mathbb{Z}$ we need to compute are in $[0; B-1]$
and outputs are in $[0; (B-1)^2 M],$ so doing calculations modulo $p$ would never result
in an ambiguity.
Note that, generally speaking, we would need to round the size of the convolution, $N,$ up to some
divisor of $(p-1)$ in order to guarantee the existence of primitive $N$-th root of unity.
We thus need to choose $p$ such that $(p-1)$ is highly composite.

Another approach is to pick $n$ pairwise distinct primes, $p_1, \ldots, p_n,$ such that
\begin{equation} \label{eq:product_inequality}
\prod\limits_{i=1}^{n} p_i > (B-1)^2 M.
\end{equation}
Then, for each $i,$ compute the convolution vector in $\mathrm{F}_{p_i}$ and use
the Chinese remainder theorem to recover the actual answer.
In this case, we would need to round the size of the convolution, $N,$ up to some divisor of
$$\gcd(p_1 - 1, \ldots, p_n - 1).$$
We thus need to choose $p_1, \ldots, p_n$ such that this value is highly composite.

Note that the latter approach is just a generalization of the first one:
just assume that $\gcd(p-1) = p-1.$

Remember also that the ``high compositeness'' of $N,$ which must divide $\gcd(p_1 - 1, \ldots, p_n - 1),$ is required
for the Cooley-Tukey algorithm to work on a length-$N$ transform. Note the coincidence!

\section{Choice of high-level algorithm}

Having chosen the number-theoretic transform, we now need to decide on the following:

\begin{itemize}
    \item what number of primes to use, and of what size (in machine words);

    \item what power of $10$ to choose as the base $B;$

    \item how to find primes with the qualities we desire.
\end{itemize}

\subsection{Choosing the size of primes in machine words}

All real-life hardware platforms have the notion of machine word; typically, we can natively add and
subtract machine words, and get the product of two machine words as a two-word value. The size of
the pointer is also typically limited to the machine word.
Suppose the machine word is $w$ bits long; define $\mu = 2^w.$

Let us fix, for every $\ell \ge 1,$ some representation in memory for elements of $\mathrm{F}_p,$
where $p$ is $\ell$ words long prime; this means that $\mu^{\ell - 1} < p < \mu^\ell.$
We then define the following functions:
\begin{itemize}
    \item $C_{\mathrm{Add}}(\ell),$ the cost of addition of two elements in $\mathrm{F}_p$ for $\ell$-word $p;$
    \item $C_{\mathrm{Sub}}(\ell),$ the cost of subtraction of two elements in $\mathrm{F}_p$ for $\ell$-word $p;$
    \item $C_{\mathrm{Mul}}(\ell),$ the cost of multiplication of two elements in $\mathrm{F}_p$ for $\ell$-word $p.$
\end{itemize}

Let us assume the following:
\begin{itemize}

    \item $\ell \cdot C_{\mathrm{Add}}(1) \le C_{\mathrm{Add}}(\ell).$
      Indeed, it should be impossible to add (subtract, compare) $\ell$-word numbers faster than
      doing $\ell$ single-word additions (subtractions, comparisons).
      Note that addition modulo $p$ is normally implemented as simple addition, comparison with
      $p$ (assuming no overflow) and conditional subtraction.

    \item $\ell \cdot C_{\mathrm{Sub}}(1) \le C_{\mathrm{Sub}}(\ell).$
      Similar to the above: subtraction modulo $p$ is normally implemented as simple subtraction,
      underflow test and conditional addition.

    \item $\ell \cdot C_{\mathrm{Mul}}(1) < C_{\mathrm{Mul}}(\ell)$ for $\ell > 1.$
      Assume $\ell$ is small enough that the optimal way to multiply two $\ell$-word numbers is the
      ``dumb'' quadratic algorithm.
      Even not considering the cost of reduction modulo $p,$
      the cost of ``raw'' multiplication of two $\ell$-word numbers into $2\ell$-word number is
      $\ell^2$ single-word multiplications with some additions, as opposed to just $\ell$
      single-word multiplications that the left-hand side of this inequality attempts to express.

\end{itemize}

Fix then a fast Fourier transform algorithm.
For a transform size of $N,$ it does $\Upsilon_{\mathrm{Add}}(N)$ additions, $\Upsilon_{\mathrm{Sub}}(N)$ subtractions, and $\Upsilon_{\mathrm{Mul}}(N)$
multiplications over $\mathrm{F}_p.$ We assume it does no divisions in $\mathrm{F}_p,$ which is a reasonable assumption if we have inverses in $\mathrm{F}_p$
to all possible values of $N$ pre-calculated.

We now want to compare the approach of using $\ell > 1$ pairwise distinct single-word primes against the approach of using a single $\ell$-word prime.
By our assumptions,
   $$\ell \big( C_{\mathrm{Add}}(1) \Upsilon_{\mathrm{Add}}(N) + C_{\mathrm{Sub}}(1) \Upsilon_{\mathrm{Sub}}(N) + C_{\mathrm{Mul}}(1) \Upsilon_{\mathrm{Mul}}(N) \big)$$
is less than
   $$C_{\mathrm{Add}}(\ell) \Upsilon_{\mathrm{Add}}(N) + C_{\mathrm{Sub}}(\ell) \Upsilon_{\mathrm{Sub}}(N) + C_{\mathrm{Mul}}(\ell) \Upsilon_{\mathrm{Mul}}(N),$$
if $\Upsilon_{\mathrm{Mul}}(N) > 0.$

This means that using $\ell$ distinct single-word primes is better;
the same argument can be invoked to show that using an ensemble of
primes of mixed word lengths is suboptimal compared to an ensemble of single-word primes.

Note that we do not consider the cost of answer recovery here,
because the transform part is $\Theta(N \log N)$
and the answer recovery part is $\Theta(N);$
and, in practice, the transform part dominates.

Neither does our analysis consider cache efficiency, which ought to be better for $\ell$ distinct
single-word primes.

\subsection{Choosing the number of primes}

Let us now choose the number of primes, $\ell.$
Remember we defined $\mu = 2^w,$ where $w$ is the length of the machine word in bits.
Remember also that by~\eqref{eq:product_inequality},
we want to pick $p_1, \ldots, p_\ell$ such that
$$\prod\limits_{i=1}^{\ell} p_i > (B-1)^2 M,$$
where $M$ is the maximum length of a multiplicand possible.

We will require
\begin{equation} \label{eq:pi_bound}
p_i < \mu/2
\end{equation}
for simpler implementation of addition modulo $p_i.$
If this does not hold, then, during addition of two numbers modulo $p_i,$ the raw sum may overflow
the machine word, and the implementation would need to check for two conditions instead of just
one: we would need to subtract $p_i$ (modulo $\mu$) from the result if either the overflow happened
or the result is greater or equal to $p_i.$

Having fixed $\mu = 2^w$ and $M,$ and assuming all $p_1, \ldots, p_\ell$ will be
approximately equal to $\mu/2,$ we can define the following function:
\begin{equation} \label{eq:lambda}
\lambda(\ell) = \max\{n \in \mathbb{N} : ({10}^n - 1)^2 < (\mu / 2)^\ell / M\}.
\end{equation}
Then, once we choose $\ell,$ we calculate $\lambda(\ell)$ and, assuming it is positive, we can put
$B = {10}^{\lambda(\ell)}.$ If $\lambda(\ell) = 0,$ the chosen value of $\ell$ is too small; more
prime moduli are needed.

Note that we do not require $B \le \mu$ here; this may seem strange, but this
only impacts initialization and answer recovery stages, which are $\Theta(N),$ and may potentially
speed up the transform stage, which is $\Theta(N \log N).$

Let us calculate the values of $\lambda(\ell)$ for $\mu \in \{2^{32}, 2^{64}\},$ $M \in \{2^{15}, 2^{20}, 2^{25}\}$ and $\ell \in \{1,2,3,4\}.$

Note the meaning of $\lambda(\ell) / \ell$: if a multiplicand has $n$ digits in base ten, it will
have $\lceil n / \lambda(\ell) \rceil$ digits in base $B = 10^{\lambda(\ell)};$ but we will need to perform $\ell$
transforms: one for each prime modulo. The ratio $\lambda(\ell) / \ell,$ thus, is approximately the
``speedup factor'' over doing a single transform with $B = 10.$

First, for $\mu = 2^{64}$:

\medskip

\begin{tabular}{ lll }

\begin{tabular}{ |c|c|c| }
 \hline
 \multicolumn{3}{|c|}{ $M=2^{15}$ } \\
 \hline\hline
  $\ell$ & $\lambda(\ell)$ & $\lambda(\ell) / \ell$ \\
 \hline\hline
  $1$ & $7$ & $7.0000$ \\
 \hline
  $2$ & $16$ & $8.0000$ \\
 \hline
  $3$ & $26$ & $8.6667$ \\
 \hline
  $4$ & $35$ & $8.7500$ \\
 \hline
\end{tabular}
&
\begin{tabular}{ |c|c|c| }
 \hline
 \multicolumn{3}{|c|}{ $M=2^{20}$ } \\
 \hline\hline
  $\ell$ & $\lambda(\ell)$ & $\lambda(\ell) / \ell$ \\
 \hline\hline
  $1$ & $6$ & $6.0000$ \\
 \hline
  $2$ & $15$ & $7.5000$ \\
 \hline
  $3$ & $25$ & $8.3333$ \\
 \hline
  $4$ & $34$ & $8.5000$ \\
 \hline
\end{tabular}
&
\begin{tabular}{ |c|c|c| }
 \hline
 \multicolumn{3}{|c|}{ $M=2^{25}$ } \\
 \hline\hline
  $\ell$ & $\lambda(\ell)$ & $\lambda(\ell) / \ell$ \\
 \hline\hline
  $1$ & $5$ & $5.0000$ \\
 \hline
  $2$ & $15$ & $7.5000$ \\
 \hline
  $3$ & $24$ & $8.0000$ \\
 \hline
  $4$ & $34$ & $8.5000$ \\
 \hline
\end{tabular}
\end{tabular}

\medskip

Then, for $\mu = 2^{32}$:

\medskip

\begin{tabular}{ lll }

\begin{tabular}{ |c|c|c| }
 \hline
 \multicolumn{3}{|c|}{ $M=2^{15}$ } \\
 \hline\hline
  $\ell$ & $\lambda(\ell)$ & $\lambda(\ell) / \ell$ \\
 \hline\hline
  $1$ & $2$ & $2.0000$ \\
 \hline
  $2$ & $7$ & $3.5000$ \\
 \hline
  $3$ & $11$ & $3.6667$ \\
 \hline
  $4$ & $16$ & $4.0000$ \\
 \hline
\end{tabular}
&
\begin{tabular}{ |c|c|c| }
 \hline
 \multicolumn{3}{|c|}{ $M=2^{20}$ } \\
 \hline\hline
  $\ell$ & $\lambda(\ell)$ & $\lambda(\ell) / \ell$ \\
 \hline\hline
  $1$ & $1$ & $1.0000$ \\
 \hline
  $2$ & $6$ & $3.0000$ \\
 \hline
  $3$ & $10$ & $3.3333$ \\
 \hline
  $4$ & $15$ & $3.7500$ \\
 \hline
\end{tabular}
&
\begin{tabular}{ |c|c|c| }
 \hline
 \multicolumn{3}{|c|}{ $M=2^{25}$ } \\
 \hline\hline
  $\ell$ & $\lambda(\ell)$ & $\lambda(\ell) / \ell$ \\
 \hline\hline
  $1$ & $0$ & $0$ \\
 \hline
  $2$ & $5$ & $2.5000$ \\
 \hline
  $3$ & $10$ & $3.3333$ \\
 \hline
  $4$ & $14$ & $3.5000$ \\
 \hline
\end{tabular}
\end{tabular}

\medskip

First, note the phenomenon of diminishing returns: consider, for example, $\mu = 2^{64}, M=2^{20};$
the speedup factor of using $\ell=2$ over $\ell=1$ is $7.5 / 6 = 1.25,$
while the speedup factor of $\ell=3$ over $\ell=2$ is $8.3333 / 7.5 = 1.1111;$
and the speedup factor of $\ell=4$ over $\ell=3$ is $8.5 / 8.3333 = 1.02.$

At the same time, the larger $\ell$ is, the higher the costs of initialization and answer recovery
are.

Note also that, starting with $\ell=3,$ for any $2^{15} \le M \le 2^{25}$ and both $\mu=2^{32}$ and $\mu=2^{64},$
the value of $B = 10^{\lambda(\ell)}$ becomes greater than $\mu;$ this means that we can not represent a value
modulo $B$ with one machine word, which means more overhead (in both performance and complexity of the code) on
initialization and answer recovery.

We thus think it is wise to choose $\ell=2$ as a nice trade-off between the performance of the
transform, costs of initialization and answer recovery, and complexity of the code.

Note that the \textbf{mpdecimal} library uses $\ell = 3$ prime moduli with bases
$B=10^{19}$ on 64-bit systems and $B=10^{9}$ on 32-bit systems; the speedup factors are
$19/3=6.3333$ on 64-bit systems and $9/3=3$ on 32-bit systems.

\subsection{Choosing the base}
\label{section:bases}

Definition~\eqref{eq:lambda} gives us a way to calculate
$B = 10^{\lambda(\ell)}$ given the values of $\mu,$ $M,$ and $\ell.$
In practice, it is better to support multiple bases, each for its own range of transform lengths.
This eliminates the need for picking the single maximum transform length $M$ and slowing down
smaller transforms.

Our implementation supports bases
$\{ 10^{14}, 10^{15}, 10^{16}, 10^{17} \}$ on 64-bit systems; and bases
$\{ 10^{5}, 10^{6}, 10^{7} \}$ on 32-bit systems.

\subsection{Choosing the method of modular reduction}

The most computationally expensive low-level operation that is carried out during the
number-theoretic transform is modular multiplication. In order to multiply two elements of
$\mathrm{F}_p,$ it is not enough to simply compute the raw product of values in $[0; p-1]$:
we need to perform reduction modulo $p;$ although, as we will see, what it exactly means depends on
the representation of elements.

We discuss the methods of reduction now because one method requires prime moduli of special form;
thus, our choice may affect our strategy of searching primes.

\subsubsection{The naïve approach}

The naïve method of modular reduction is to use, where it is available, a hardware instruction
to divide two-words dividend by single-word divisor into single-word quotient and single-word
remainder. Where not available, we would have to emulate such an instruction in software.

Note that such an instruction exists in x86 and x86-64.
As for software emulation, both GNU GCC and Clang compilers provide \verb!uint64_t! type on 32-bit
platforms, and \verb!unsigned __int128! type on 64-bit platforms,
with support for the division operation.

Unfortunately, this method is very slow; see section~\ref{section:modmul_bench} for the
results of our benchmark against Montgomery reduction and Solinas reduction.

\subsubsection{Barrett reduction}

A method intended to be faster, while not requiring a change in our representation of field
elements, is known as Barrett reduction~\cite{barrett}.

It reduces $0 \le a < n^2$ modulo $n$ using some pre-computed value that depends on $n.$
For a $k$-bit modulo $n,$ it internally (not counting the raw multiplication $xy = a$) performs:
\begin{itemize}
    \item one multiplication of $(k+1)$ bits by $(2k)$ bits into a $(3k)$-bit value;
    \item one multiplication of $k$ bits by $k$ bits, of which only the lowest $k$ bits are used.
\end{itemize}

\subsubsection{Montgomery reduction}
\label{section:montgomery_intro}

A method even faster for our purposes, is Montgomery reduction~\cite{Montgomery1985ModularMW}.

Remember we defined $w$ as the bit width of the machine word;
define then
\begin{equation} \label{eq:mont_r}
R = 2^w \in \mathrm{F}_p.
\end{equation}
We assume $p > 2,$ so $R$ is non-zero in $\mathrm{F}_p.$

Then, the \textit{Montgomery representation} of $x \in \mathrm{F}_p$ is simply $Rx.$

If $Rx$ is the Montgomery representation of $x,$ and $Ry$ is the Montgomery representation of $y,$
then the Montgomery representation of $(x \pm y)$ is simply $(Rx \pm Ry)$ as $R(x \pm y) = Rx \pm Ry;$
it means that Montgomery representations can be added and subtracted as ordinary values modulo $p.$

The \textit{Montgomery reduction} is a function
$\redc \colon \mathrm{F}_p \times \mathrm{F}_p \to \mathrm{F}_p$
defined as follows:
\begin{equation} \label{eq:mont_redc}
\redc(x, y) = R^{-1} x y.
\end{equation}
It is important because the Montgomery representation $R(x y)$ of the product $x y$ is
$\redc(Rx, Ry).$ Also, any $x \in \mathrm{F}_p$ can be converted into Montgomery
representation by invoking $\redc(x, R^2) = Rx;$ and out of Montgomery
representation by invoking $\redc(x, 1) = R^{-1}x.$

It is possible to compute $\redc$ efficiently. We now give the definition of
the procedure $\mathrm{MontgomeryReduce}(a).$
If we represent field elements as values in $[0; p-1],$ then
$\redc(x,y)$ can be computed as $\mathrm{MontgomeryReduce}(xy).$

\begin{procedure}[!hbt]
\DontPrintSemicolon
\caption{MontgomeryReduce($a$)}
\KwData{
    Integer $\mu.$\;
    Integer $p$ coprime with $\mu.$\;
    Integer $p' \in [0; \mu-1]$ such that $p \cdot p' \equiv -1 \, (\mymod \, \mu).$
}
\KwIn{
    Integer $a \in [0; \mu p - 1].$
}
\KwOut{
    Integer $b \in [0; p-1]$ such that $b \equiv a \mu^{-1} \, (\mymod \, p).$
}
\Begin{
    $m \longleftarrow ((a \mymod \mu) \cdot p') \mymod \mu$\;
    $r \longleftarrow (a + m \cdot p) \mydiv \mu$
        \tcc*[r]{this division is exact} 
    \eIf{$r \ge p$}{
        \Return{$r - p$}\;
    }{
        \Return{$r$}\;
    }
}
\end{procedure}

For the pre-calculated data, put $\mu = 2^w,$ $p > 2$ is our prime number, and
the value of $p'$ can be calculated with the extended Euclidean algorithm for GCD;
we have $\gcd(p, \mu) = 1,$ so $\exists c_1, c_2 \in \mathbb{Z} \colon c_1 p + c_2 \mu = 1.$
This means $c_1 p \equiv 1 \, (\mymod \, \mu)$ and $-c_1 p \equiv -1 \, (\mymod \, \mu),$ so
$$p' = (-c_1) \mymod \mu.$$

The procedure $\mathrm{MontgomeryReduce(a)}$ internally
(not counting the raw multiplication $xy = a$)
performs:
\begin{itemize}
    \item one multiplications of $w$ bits by $w$ bits into $(2w)$ bits;
    \item one multiplications of $w$ bits by $w$, of which only the lowest $w$ bits are used.
\end{itemize}
This is more efficient than the Barrett reduction if one does not consider the cost of conversion
into and out of Montgomery representation. In the case of number-theoretic transform, we can
convert everything into Montgomery representations at the beginning (this is $\Theta(N)$ time),
then perform the transform (this is $\Theta(N \log N)$ modular multiplications), and then
convert everything out of Montgomery representation (this is, again, $\Theta(N)$ time).

Better still, we can completely omit those conversions, instead mixing a factor into the final stage
of multiplication by $N^{-1}$ in the inverse transform; see section~\ref{section:trick} for details.

\subsubsection{Solinas primes}

The \textbf{mpdecimal} library uses yet another method of modular reduction on 64-bit systems;
it is based on the use of prime moduli of form $2^{64} - 2^{n} + 1.$
Generally, such primes are known as Solinas primes, or generalized Mersenne
primes~\cite{Solinas},
defined as primes of form $f(2^m),$ where $f(x)$ is a low-degree polynomial with small integer
coefficients.

The single round of reduction modulo $p = 2^{64} - 2^{n} + 1$ is then defined as follows:
$$r(2^{64} x_1 + x_0) = 2^{n} x_1 - x_1 + x_0,$$
where $x_0, x_1 \in \mathbb{N}$ and $x_0, x_1 < 2^{64}.$
After some small number of rounds (2---3 rounds for primes used in \textbf{mpdecimal}), the result
is guaranteed to be less than $2p,$ after which it can be reduced modulo $p$ with a single
conditional subtraction, just like with the Montgomery reduction.

We benchmarked the ``best case'' of reduction modulo $2^{N_1}-2^{N_2}+1$ against Montgomery
reduction: we used $2^{64} - 2^{32} + 1$ as the prime modulo; note the values of $64$ and
$32$ are better for hardware (at least, for x86-64) because division and multiplication can be done
by simply omitting (half-)words or assuming zero lower (half-)words, correspondingly, instead of
actually shifting the bits. This prime also requires at most 2 rounds of reduction.

We found that on a modern x86-64 system, reduction with two rounds of $r$ is slower by
$\approx$10\% compared to Montgomery reduction;
see section~\ref{section:modmul_bench}.

We think the most likely explanation for this finding is that, on modern x86-64 systems, hardware
multipliers are sufficiently performant to render special schemes for reduction modulo a
single-word $p=2^{N_1}-2^{N_2}+1$ useless, if Montgomery reduction can be used instead.

\subsubsection{Benchmark}
\label{section:modmul_bench}

We benchmarked naïve division using x86-64
\verb~divq~
instruction, Montgomery reduction,
and Solinas reduction.

Given an argument $n,$ our benchmark performs $n$ iterations of loop with
$10$ unrolled modular multiplications, performing in total $10 n$ modular multiplications.

We compiled it using Clang 11.0.0, with option \verb!-O3!. The machine is
Xiaomi RedmiBook 14" 2019 JYU4203CN laptop with
Intel\textsuperscript{\textregistered} Core\texttrademark{} i3-8145U CPU @ 2.10GHz;
CPU scaling governors for all CPUs were set to ``performance''.

\begin{center}
\begin{tabular}{ |c|c|c| }
    \hline
        Approach & $n$ & Time, s \\
    \hline\hline
        Naïve & $10^8$ & 23.04 \\
    \hline
        Montgomery & $10^8$ & 2.86 \\
    \hline
        Solinas & $10^8$ & 3.19 \\
    \hline
\end{tabular}
\end{center}

The code can be found in the \verb!bench-modmul! subdirectory of our repository
(see section~\ref{section:avail}).

\subsection{Searching for primes}

Having chosen the Montgomery reduction, we are going search for primes of form
\begin{equation} \label{eq:pformula}
c \cdot 3 \cdot 2^n + 1.
\end{equation}

The factor of $2^n$ here ensures that we will be able to perform transforms of length $2^m$ for any $m \le n;$
and the factor of $3$ is here to ``smooth the stairs'', allowing us to perform transforms of length
$3 \cdot 2^m$ for any $m \le n.$ Note that $3 \cdot 2^m = 1.5 \cdot 2^{m+1}$ is exactly the average
of $2^{m+1}$ and $2^{m+2}.$

Observe that, in order to guarantee the uniqueness of representation of a prime as~\eqref{eq:pformula},
we need to require $c$ to be odd --- otherwise a power of two can be factored out into $2^n.$

Define $n_{\mathrm{min}} = \lceil \log_2(M/3) \rceil,$ the lower bound for $n$ in~\eqref{eq:pformula}.
We have, then, the following requirements for each $p_i$:
\begin{enumerate}
    \item $p_i$ is of form $c \cdot 3 \cdot 2^n + 1,$ where $c$ is odd and $n \ge n_{\mathrm{min}};$
    \item $p_i < \mu / 2.$
\end{enumerate}
Of all possible primes with those properties, we need to pick the $\ell$ largest.

This leads us to the following algorithm:

\begin{procedure}[!hbt]
\caption{FindPrimesForN($n$, $p_{\max}, \ell$)}
\SetKw{KwAnd}{and}
\DontPrintSemicolon
\KwIn{$n, p_{\max}, \ell.$}
\KwOut{Set of $\ell$ largest, or less if the total number is less than $\ell,$ primes of form $p = c \cdot 3 \cdot 2^n + 1,$ where $p \le p_{\max}$ and $c$ is odd.}
\Begin{
    $r \longleftarrow \varnothing$\;
    $\psi \longleftarrow (p_{\max} - 1) \mydiv 2^n $\;
    \If{$\psi \mymod 2 = 0$}{
        $\psi \longleftarrow \psi - 1$\;
    }
    \While{$\psi \mymod 3 \ne 0$}{
        $\psi \longleftarrow \psi - 2$\;
    }
    \While{$\psi > 0$ \KwAnd $|r| < \ell$}{
        $p \longleftarrow \psi \cdot 2^n + 1$\;
        \If{$p$ is prime}{
            $r \longleftarrow r \cup \{p\}$\;
        }
        $\psi \longleftarrow \psi - 6$\;
    }
    \Return{$r$}\;
}
\end{procedure}

\begin{procedure}[!hbt]
\caption{FindPrimes($w$, $n_{\min}$, $\ell$)}
\SetKw{Error}{error}
\DontPrintSemicolon
\KwIn{$w, n_{\min}, \ell.$}
\KwOut{Set of $\ell$ largest primes of form $p = c \cdot 3 \cdot 2^n + 1,$ where $p < 2^{w-1},$ $n \ge n_{\min},$ and $c$ is odd.}
\Begin{
    $r \longleftarrow \varnothing$\;
    $p_{\max} \longleftarrow 2^{w-1}-1$\;
    \For{$n \longleftarrow n_{\min}$ \KwTo $w-2$}{
        $r \longleftarrow r \cup \mathrm{FindPrimesForN}(n, p_{\max}, \ell)$\;
    }
    \eIf{$|r| < \ell$}{
        \Error{cannot find $\ell$ primes with required properties}\;
    }{
        \Return{$\ell$ largest values of $r$}\;
    }
}
\end{procedure}

Note that it uses an unspecified algorithm to perform primality testing.
We use the Miller-Rabin primality test with the first 12 prime numbers as bases; it has been proven
in~\cite{primebases} that, for values less than $2^{64},$ this is enough to guarantee correctness.

\section{Cooley-Tukey optimizations}

Before discussing optimizations,
we need to define the decimation in time (DIT) and decimation in frequency (DIF) variants of the
Cooley-Tukey FFT algorithm.

Both are specializations of Cooley-Tukey for a power-of-two length, and both include the step of
applying the bit-reversal permutation, which we introduce below, to the array being transformed,
either at the very beginning or in the very end.

Both algorithms take two parameters, an array $A$ of field elements and a field element
$\omega = \xi^{\pm 1},$
and return an array $A'$ of length $|A'| = |A|$
such that $$A'[k] = \sum\limits_{i=0}^{|A| - 1} A[i] \omega^{ik}.$$
If we express this transform as a function $\tau(A,\omega),$ then DFT and IDFT of length $N$ can be
expressed in terms of $\tau$ as follows:
\begin{subequations}
\begin{align}
\mathfrak{f}(x) &= \tau(x, \xi); \\
\mathfrak{f}^{-1}(x) &= N^{-1} \tau(x, \xi^{-1}).
\end{align}
\end{subequations}

\subsection{Bit-reversal permutation}

Bit-reversal permutation is defined for sequences of length $2^n$ as follows: element with
zero-based index $k$ is exchanged with element with zero-based index $\mathbf{rev}_n(k),$ where
$\mathbf{rev}_n(k)$ is defined as the unique number in $[0; 2^n - 1]$ whose length-$n$ binary
representation equals to the reversed length-$n$ binary representation of $k.$

We now give the definition of the procedure that performs bit-reversal permutation.

\begin{procedure}[!hbt]
\caption{BitRevPermute($A$)}
\DontPrintSemicolon
\KwIn{Array $A$, $|A| = 2^n.$}
\KwOut{The result of applying the bit-reversal permutation to $A.$}
\Begin{
    $n \longleftarrow \log_2 |A|$\;
    \For{$k \longleftarrow 0$ \KwTo $2^n - 1$}{
        $\langle b_0, \ldots, b_{n-1} \rangle \longleftarrow$ bits of $k;$ $b_i \in \{0,1\}$ and $\sum_{i=0}^{n-1} b_i 2^i = k$\;
        $k' \longleftarrow \sum_{i=0}^{n-1} b_{n-i-1} 2^i$\;
        \If{$k < k'$}{
            {exchange $A[k]$ with $A[k']$}\;
        }
    }
    \Return{$A$}\;
}
\end{procedure}

\subsection{Decimation in time}

We now define the decimation in time (DIT) algorithm.
Note that it performs bit-reversal permutation as the first step.
$\widehat{1}$ denotes the multiplicative identity of the underlying field.

\begin{procedure}[!hbt]
\caption{DecimationInTime($A$, $\omega$)}
\SetKw{KwBy}{by}
\DontPrintSemicolon
\KwIn{
    Array $A$ of field elements, $|A| = 2^n$.
    Field element $\omega = \xi^{\pm 1},$ where $\xi$ is a primitive $2^n$-th root of unity.
}
\KwOut{
    Array $A'$ such that $|A'| = |A|, A'[k] = \sum_{i=0}^{|A| - 1} A[i] \omega^{ik}.$
}
\Begin{
    $A \longleftarrow \mathrm{BitRevPermute}(A)$\;
    \For{$s \longleftarrow 1$ \KwTo $\log_2 |A|$}{
        $m \longleftarrow 2^s$\;
        $\omega_m \longleftarrow \omega^{|A| / m}$\;
        \For{$k = 0$ \KwTo $|A|-1$ \KwBy $m$}{
            $\varphi \longleftarrow \widehat{1}$\;
            \For{$j = 0$ \KwTo $m/2-1$}{
                $u \longleftarrow A[k + j]$\;
                $v \longleftarrow \varphi \cdot A[k + j + m/2]$\;
                $A[k + j] \longleftarrow u + v$\;
                $A[k + j + m/2] \longleftarrow u - v$\;
                $\varphi \longleftarrow \varphi \cdot \omega_m$\;
            }
        }
    }
    \Return{$A$}\;
}
\end{procedure}

The four-line transform on $A[k+j]$ and $A[k+j+m/2]$ is known as ``butterfly'', or,
more specifically, this one is the Cooley-Tukey butterfly; it maps
$\langle x,y \rangle \mapsto \langle x', y' \rangle$ as follows:
\begin{subequations}
\begin{align}
x' &= x + \varphi \cdot y; \\
y' &= x - \varphi \cdot y.
\end{align}
\end{subequations}

\subsection{Decimation in frequency}

We now define the decimation in frequency (DIF) algorithm.
Note that it performs bit-reversal permutation as the last step.
As in the previous subsection, $\widehat{1}$ denotes the multiplicative identity of the underlying field.

\begin{procedure}[!hbt]
\caption{DecimationInFrequency($A$, $\omega$)}
\SetKw{KwBy}{by}
\SetKw{KwDownTo}{downto}
\DontPrintSemicolon
\KwIn{
    Array $A$ of field elements, $|A| = 2^n$.
    Field element $\omega = \xi^{\pm 1},$ where $\xi$ is a primitive $2^n$-th root of unity.
}
\KwOut{
    Array $A'$ such that $|A'| = |A|, A'[k] = \sum_{i=0}^{|A| - 1} A[i] \omega^{ik}.$
}
\Begin{
    \For{$s \longleftarrow \log_2 |A|$ \KwDownTo $1$}{
        $m \longleftarrow 2^s$\;
        $\omega_m \longleftarrow \omega^{|A| / m}$\;
        \For{$k = 0$ \KwTo $|A|-1$ \KwBy $m$}{
            $\varphi \longleftarrow \widehat{1}$\;
            \For{$j = 0$ \KwTo $m/2-1$}{
                $u \longleftarrow A[k + j]$\;
                $v \longleftarrow A[k + j + m/2]$\;
                $A[k + j] \longleftarrow u + v$\;
                $A[k + j + m/2] \longleftarrow \varphi \cdot (u - v)$\;
                $\varphi \longleftarrow \varphi \cdot \omega_m$\;
            }
        }
    }
    $A \longleftarrow \mathrm{BitRevPermute}(A)$\;
    \Return{$A$}\;
}
\end{procedure}

The four-line transform on $A[k+j]$ and $A[k+j+m/2]$ is known as the Gentleman-Sande butterfly; it maps
$\langle x,y \rangle \mapsto \langle x', y' \rangle$ as follows:
\begin{subequations}
\begin{align}
x' &= x + y; \\
y' &= \varphi \cdot (x - y).
\end{align}
\end{subequations}

\subsection{Omitting the bit-reversal permutation}
\label{section:omitting1}

Note that in~\eqref{eq:conv_theorem}, for the computation of $(\_ \star \_),$ the specific order in
which the direct transform $\mathfrak{f}$ produces its outputs, and the order in which the inverse
transform $\mathfrak{f}^{-1}$ expects its inputs to be in, do not matter; we only need these two
orders to match.
Indeed, $(\_ \cdot \_)$ is element-wise, $(x \cdot y)[i] = x[i] y[i],$
so, for any permutation $\pi$ we have
\begin{equation} \label{eq:cdot_permutation_invariance}
\pi^{-1}(\pi(x) \cdot \pi(y)) = x \cdot y.
\end{equation}

This means that, for the purpose of computing the cyclic convolution, we can use DIF for the direct
transform and DIT for the inverse transform, and simultaneously omit the final permutation step in
DIF and the initial permutation step in DIT.
Note this does not mean just one of them can be omitted but not the other.

\subsection{Pre-calculating the table of root powers}

Define $N = \log_2 |A|.$

Note that, in both DIT and DIF, we perform the same computation over and over again when doing
$$\varphi \longleftarrow \varphi \cdot \omega_m,$$
where $\omega_m = \omega^{2^{N - s}}, s = \log_2 m.$

One way to improve that is to pre-calculate powers of $\omega$ into a table $t,$ so that
$t[i] = \omega^i, 0 \le i < |A|/2.$
Then, given $j$ and $m=2^s,$ we can express the factor $\varphi$ as $t[j \cdot 2^{N - s}].$
Note that the multiplication by a power of two can be replaced with a bit shift.

Another way is to pre-calculate a separate table for each ``granularity'' of root powers:
define $$T_s[i] = t[i \cdot 2^{N-s}] = (\omega^{2^{N-s}})^i, 0 \le i < 2^{s-1}$$
for each $s, 1 \le s \le N.$ Note $T_N = t$ and $\sum\limits_{s=1}^{N} |T_s| = \sum\limits_{s=1}^{N} 2^{s-1} = 2^N-1.$
Now, given $j$ and $m=2^s,$ we can express $\varphi$ as $T_s[j].$

Let us write $T_s$ for all $s$ into a single array $T_\ast$
of length $2^N-1 = |A|-1$ so that
$$T_s[i] = T_\ast[(2^0 + 2^1 + \cdots + 2^{s-2}) + i] = T_\ast[2^{s-1} - 1 + i].$$
Then, given $j$ and $m=2^s,$ the factor $\varphi$ can be expressed as
$T_\ast[\delta + j],$ where $\delta = 2^{s-1}-1 = m/2 - 1.$

At first sight, there is not much difference between indexing
$T_\ast[C_1 + j]$ and $t[j \cdot 2^{C_2}]$
in the innermost loop,
where $C_1$ and $C_2$ denote loop-invariant expressions that can be calculated once before the loop.
But on x86 and x86-64 platforms, there \textit{is} a difference:
they can encode SIB (scale-index-base) directly in the instruction that dereferences the
pointer, but only for constant scales of 1, 2, 4 and 8 bytes.
This means that, if our field element is 8, 16, 32 or 64 bits long, we can dereference $p[j],$
where $p$ points to $T_\ast[C_1],$ in one instruction; but dereferencing
$t[j \cdot 2^{C_2}],$ where $C_2$ can be anything, requires two instructions: first for the bit
shift and the second for dereferencing.

\subsection{Separate factor-1 butterfly}

Note that, in both DIT and DIF, the factor $\varphi$ for the step $j=0$ is $\widehat{1},$
the multiplicative identity of the field.
In this case, both butterflies degenerate into
$\langle x, y \rangle \mapsto \langle x+y, x-y \rangle.$

We can peel the first iteration off the loop to avoid redundant multiplication by $\widehat{1}.$

\section{The linearity trick}
\label{section:trick}

We are now returning to the Montgomery reduction; we defined $R$ and $\redc$
in~\ref{section:montgomery_intro}; for the purposes of this section, $p > 2$ is arbitrary prime.

Consider the field $\widetilde{\mathrm{F}}_p,$ which is the same as $\mathrm{F}_p,$ but
having $\redc$ as the multiplication operation.
It is a field isomorphic to $\mathrm{F}_p,$ with isomorphism
$\varphi \colon \mathrm{F}_p \to \widetilde{\mathrm{F}}_p$ being $\varphi(x) = Rx.$

If we fix a primitive $N$-th root of unity $\xi \in \mathrm{F}_p,$
then $\widetilde{\xi} = R \xi$ will be a primitive $N$-th root of unity
in $\widetilde{\mathrm{F}}_p.$

We can consider, then,
$\widetilde{\mathfrak{f}}$ and $\widetilde{\mathfrak{f}}^{-1},$
the DFT and IDFT, correspondingly, for the field $\widetilde{\mathrm{F}}_p.$
We have the following identities, by the isomorphism, for any $x \in \mathrm{F}_p^N$:
\begin{subequations}
\begin{align}
    \mathfrak{f}(x)         &= R^{-1} \, \widetilde{\mathfrak{f}}(R \, x); \\
    \mathfrak{f}^{-1}(x)    &= R^{-1} \, \widetilde{\mathfrak{f}}^{-1}(R \, x).
\end{align}
\end{subequations}

Rewrite with $x = R^{-1}y$:
\begin{subequations}
\begin{align}
    \mathfrak{f}(R^{-1}y)       &= R^{-1} \, \widetilde{\mathfrak{f}}(y); \\
    \mathfrak{f}^{-1}(R^{-1}y)  &= R^{-1} \, \widetilde{\mathfrak{f}}^{-1}(y).
\end{align}
\end{subequations}

Rewrite the left-hand sides by the linearity of $\mathfrak{f}$ and $\mathfrak{f}^{-1}$
(see~\eqref{eq:transform_linearity}):
\begin{subequations}
\begin{align}
    R^{-1} \mathfrak{f}(y)       &= R^{-1} \, \widetilde{\mathfrak{f}}(y); \\
    R^{-1} \mathfrak{f}^{-1}(y)  &= R^{-1} \, \widetilde{\mathfrak{f}}^{-1}(y).
\end{align}
\end{subequations}

Dividing both sides by $R^{-1},$ we see that the transforms in $\mathrm{F}_p$ and
$\widetilde{\mathrm{F}}_p$ completely coincide: for any
$y \in \mathrm{F}_p^N,$
\begin{subequations} \label{eq:dft_idft_in_tildefp}
\begin{align}
\mathfrak{f}(y) &= \widetilde{\mathfrak{f}}(y); \\
\mathfrak{f}^{-1}(y) &= \widetilde{\mathfrak{f}}^{-1}(y).
\end{align}
\end{subequations}

Let us now consider the cyclic convolution $\widetilde{\star}$ for $\widetilde{\mathrm{F}}_p;$
by~\eqref{eq:conv_theorem}, we know that
$$x \, \widetilde{\star} \, y = \widetilde{\mathfrak{f}}^{-1}
\big( \thinspace \widetilde{\mathfrak{f}}(x) \, \widetilde{\otimes} \, \widetilde{\mathfrak{f}}(y) \big),$$
where $\widetilde{\otimes}$ denotes element-wise $\redc$
(which is, remember, the multiplication operation of $\widetilde{\mathrm{F}}_p.$)
Having proven~\eqref{eq:dft_idft_in_tildefp}, we can rewrite it as follows:
$$x \, \widetilde{\star} \, y = R^{-1} \mathfrak{f}^{-1}\big( \mathfrak{f}(x) \cdot \mathfrak{f}(y) \big),$$
where $\cdot$ denotes element-wise product.
This, in part, can be rewritten as
$$x \, \widetilde{\star} \, y = R^{-1}(x \star y).$$

It means that, computing a cyclic convolution via DFT of both arguments, followed by element-wise
multiplication, followed by IDFT, but using
$\redc$ instead of regular multiplication during all of those operations,
gives us an extra factor of $R^{-1}.$

Instead of converting both arguments into Montgomery representation
(remember this conversion is just multiplication by $R$), performing $\widetilde{\star}$ and
converting the result out of Montgomery representation (this is multiplication by $R^{-1}$),
we can just convert one of the arguments; indeed,
$$x \, \widetilde{\star} \, (Ry) = R^{-1}(x \star Ry) = x \star y.$$

Alternatively, we can multiply the result by $R,$ not touching any of the arguments at all:
$$R(x \, \widetilde{\star} \, y) = R R^{-1}(x \star y) = x \star y.$$

The latter approach is preferable for us: the Cooley-Tukey IDFT performs multiplication by
scalar $N^{-1}$ as a separate final stage; we can ``mix in'' the factor of $R$ into this stage
essentially for free.

Note that multiplication by $R$ is $\redc$ with $R^2;$ so if the Montgomery representation of
the factor $N^{-1}$ was $\psi = RN^{-1},$
then our new factor will be
$$\psi' = \redc(\psi, R^2) = R^2 N^{-1}.$$
Then, $\forall x \in \mathrm{F}_p \colon \redc(x, \psi') = R N^{-1} x,$ exactly what we need.

Another practical advantage of modification of the result as opposed to one of the arguments is
that, it might be possible that we are computing the cyclic convolution of the sequence with
\textit{itself}: the sequences have not \textit{just} identical values, but
\textit{the same location in memory}; so multiplying one by $R$ automatically multiplies another.
Of course, whether such a situation is allowed depends on the implementation; but we do support
such a use case.

\section{The matrix Fourier algorithms}

We now describe the four-step and six-step algorithms for DFT/IDFT, which are matrix Fourier
algorithms, meaning that they interpret sequences of length $N = N_1 N_2$ as $N_1 \times N_2$
matrices~\cite[p. 438--439]{mattcomp}.

We assume that rows, as opposed to columns, occupy contiguous ranges in underlying sequences;
in other words, we assume row-major order:
$$
\langle 1, 2, 3, 4, 5, 6, 7, 8 \rangle \longleftrightarrow
\begin{bmatrix}
1 & 2 & 3 & 4 \\
5 & 6 & 7 & 8
\end{bmatrix}.
$$

\subsection{The four-step algorithm}

For the purposes of this subsection:
\begin{itemize}

    \item $[N_1 \times N_2]$ means ``interpreting the sequence as a matrix with $N_1$ rows and $N_2$ columns''.

    \item If a matrix has $N_1$ rows and $N_2$ columns, then its element with
        row index $i,$ where $0 \le i < N_1,$
        and column index $j,$ where $0 \le j < N_2,$ is said to be indexed $\langle i, j \rangle.$

    \item IDFT* means IDFT without multiplication by $N^{-1}.$

\end{itemize}

Fix the length of transform $N.$
Fix also $\xi_N,$ a primitive $N$-th root of unity in the underlying field.

DFT of length $N = RC$ can be computed as follows:
\begin{enumerate}
    \item $[R \times C]$, perform (length $R$) DFT on each column with $\xi = (\xi_N)^C.$
    \item $[R \times C]$, multiply each element indexed $\langle i,j \rangle$ by $(\xi_N)^{ij}.$
    \item $[R \times C]$, perform (length $C$) DFT on each row with $\xi = (\xi_N)^R.$
    \item $[R \times C]$, transpose the matrix.
\end{enumerate}

IDFT* of length $N = RC$ can be computed as follows:
\begin{enumerate}
    \item $[C \times R],$ transpose the matrix.
    \item $[R \times C],$ perform (length $C$) IDFT* on each row with $\xi = (\xi_N)^R.$
    \item $[R \times C],$ multiply each element indexed $\langle i,j \rangle$ by $(\xi_N)^{-ij}.$
    \item $[R \times C],$ perform (length $R$) IDFT* on each column with $\xi = (\xi_N)^C.$
\end{enumerate}

\subsection{The six-step algorithm}

For the purposes of this subsection:
\begin{itemize}

    \item $[N_1 \times N_2]$ means ``interpreting the sequence as a matrix with $N_1$ rows and $N_2$ columns''.

    \item If a matrix has $N_1$ rows and $N_2$ columns, then its element with
        row index $i,$ where $0 \le i < N_1,$
        and column index $j,$ where $0 \le j < N_2,$ is said to be indexed $\langle i, j \rangle.$

    \item IDFT* means IDFT without multiplication by $N^{-1}.$

\end{itemize}

Fix the length of transform $N.$
Fix also $\xi_N,$ a primitive $N$-th root of unity in the underlying field.

DFT of length $N = RC$ can be computed as follows:
\begin{enumerate}
    \item $[R \times C],$ transpose the matrix.
    \item $[C \times R],$ perform (length $R$) DFT on each row with $\xi = (\xi_N)^C.$
    \item $[C \times R],$ transpose the matrix.
    \item $[R \times C],$ multiply each element indexed $\langle i,j \rangle$ by $(\xi_N)^{ij}.$
    \item $[R \times C],$ perform (length $C$) DFT on each row with $\xi = (\xi_N)^R.$
    \item $[R \times C],$ transpose the matrix.
\end{enumerate}

IDFT* of length $N = RC$ can be computed as follows:
\begin{enumerate}
    \item $[C \times R],$ transpose the matrix.
    \item $[R \times C],$ perform (length $C$) IDFT* on each row with $\xi = (\xi_N)^R.$
    \item $[R \times C],$ multiply each element indexed $\langle i,j \rangle$ by $(\xi_N)^{-ij}.$
    \item $[R \times C],$ transpose the matrix.
    \item $[C \times R],$ perform (length $R$) IDFT* on each row with $\xi = (\xi_N)^C.$
    \item $[C \times R],$ transpose the matrix.
\end{enumerate}

\subsection{Discussion}

If $\xi_N$ is a primitive $N$-th root of unity in a field $\mathcal{F},$
and $N = N_1 N_2,$ then $\xi_{N_1} = (\xi_N)^{N_2}$ is a primitive $N_1$-th root of unity in
$\mathcal{F}.$
This means that we can indeed perform DFTs/IDFTs on rows/columns with the ``custom'' primitive
roots specified in the descriptions of the algorithms.

The four-step algorithm is just a restatement, in terms of $N_1 \times N_2$ matrix, of the
Cooley-Tukey algorithm that, in its general form, re-expresses a transform of length $N = N_1 N_2$
in terms of smaller transforms of lengths $N_1$ and $N_2.$
Thus, it can be used when it is desirable to perform a transform of a length that is not a
power of two, so that the decimation in time (DIT) or decimation in frequency (DIF) variants of the
Cooley-Tukey algorithm can not be employed directly.

The six-step algorithm has the advantage that, aside from matrix transpositions, it only accesses
memory in strides of $R$ and $C.$ Thus, it might be faster in settings of hierarchical memory,
including external-memory transforms and modern systems with multiple levels of cache.

\subsection{Omitting matrix transpositions}

Remember in section~\ref{section:omitting1} we established~\eqref{eq:cdot_permutation_invariance}.
We then used this property to get rid of bit-reversal permutations in DIT and DIF.
It turns out that, for both four-step and six-step algorithms,
the same property can be employed to get rid of the
final transposition step in DFT and the initial transposition step in IDFT* simultaneously.

For the four-step algorithm, this means no bit-reversal permutation or matrix transposition is ever
needed: we can also simultaneously omit the bit-reversal permutation steps in DIF and DIT
if we use DIF for the direct transforms in definition of four-step DFT, and DIT for inverse
transforms in the definition of four-step IDFT*.

For the six-step algorithm, unfortunately, things are not so simple.
We can simultaneously omit the final transposition step (step 6) in the direct transform
and the initial transposition step (step 1) in the inverse transform; and can use DIF without
bit-reversal permutations for step 5 of the direct transform and DIT without bit-reversal
permutations for step 2 of the inverse transform. But we still have two matrix transposition steps
per transform, and no matter whether we use DIT or DIF for step 2 of the direct transform
and step 5 of the inverse transform, we need to perform bit-reversal permutation.

\subsection{Overview of matrix transposition}

We will now discuss matrix transposition.
Matrix transposition can be done either out-of-place or in-place.

Out-of-place transposition means a separate array in the desired order is produced; this means
using $O(N_1 N_2)$ additional memory, where $N_1$ and $N_2$ are dimensions of the matrix.

The definitions of in-place transposition vary, but generally it is defined as an algorithm that
uses ``much less'' additional memory than $O(N_1 N_2).$

Another property that we might want from a transposition algorithm is cache friendliness.

Note that a cache-oblivious algorithm for out-of-place matrix transposition is
well-known~\cite{cache_rect_mat_trans}.
We now want to explore the space of in-place transposition algorithms.

The algorithm for in-place transposition of a square matrix is simple: element
$\langle i, j \rangle$ is exchanged with element $\langle j, i \rangle.$
The same approach works for transposing a square submatrix of a larger, generally rectangular
matrix.
For completeness, we provide the code of $\mathrm{TransposeSquareSubMatrix}(p, n, n').$
For a cache-oblivious version of this algorithm, see~\cite{cache_square_mat_trans}.

\begin{procedure}[!hbt]
\caption{TransposeSquareSubMatrix($p$, $n$, $n'$)}
\DontPrintSemicolon
\KwData{
    Pointer $p$ to the first row, first column of $n$-by-$n$ submatrix of a possibly larger matrix
    with $n''$ columns and $n'$ rows stored in row-major order.
    (Formally, if the submatrix starts at $i$-th row, $0 \le i < n'$ and
    $j$-th column, $0 \le j < n'',$ then $p = q + n' \cdot i + j,$ where $q$ is the pointer to the
    beginning of the larger matrix.)

    Integer $n.$

    Integer $n'.$
}
\KwResult{
    The square submatrix is transposed.
}
\Begin{
    \For{$i = 0$ \KwTo $n-1$}{
        \For{$j = i + 1$ \KwTo $n-1$}{
            {exchange $p[i \cdot n' + j]$ with $p[j \cdot n' + i]$}\;
        }
    }
}
\end{procedure}

We will now discuss in-place transposition of a rectangular matrix.

In its general form, in-place transposition of a non-square matrix is quite hard.
The classical study begins with defining the function $\mathcal{P}(a)$ such that,
when transposing a matrix with $N_1$ rows and $N_2$ columns in row-major order,
the element with index $a$ is sent to index $\mathcal{P}(a).$
It can be defined explicitly as follows:
\begin{equation}
\mathcal{P}(a) = \begin{cases}
    N_1 N_2 - 1                 & \text{if } a = N_1 N_2 - 1; \\
    N_1 a \mymod (N_1 N_2 - 1)  & \text{otherwise.}
\end{cases}
\end{equation}
There is, then, a result~\cite{fancything} saying that
the number of fixed points of $\mathcal{P}$ is exactly
$$
1 + \gcd(N_1-1, N_2-1);
$$
and the number of cycles of length $k>1$ of $\mathcal{P}$ is
$$
\frac{1}{k} \sum\limits_{d | k} \widetilde{\mu}(k/d) \gcd(N_1^d - 1, N_1 N_2 - 1),
$$
where $\widetilde{\mu}$ is the Möbius function and the summation is performed over all divisors of
$k.$

Most algorithms for in-place transposition are essentially of ``follow-the-cycles'' kind.
This means that they iterate over all the cycles, and for each cycle,
they shift its elements cyclically. There are, then, various approaches for
locating the cycles, identifying the first element of a cycle, and tracking which cycles were
already visited.
If we limit ourselves to using only $O(N_1+N_2)$ additional memory,
even more complex algorithms are generally needed;
\cite{Fich95permutingin}
presents an algorithm with worst-case complexity of $O(N_1 N_2 \log(N_1 N_2)).$

Notably, the algorithm for matrix transposition in \textbf{mpdecimal} is also
``follow-the-cycles'' in nature, requiring $O(N_1 N_2)$ additional memory, although with a
small constant: it uses a bit set to track which elements were already shifted.

In the circumstances described above, one would believe,
we have no choice but to give up and accept the complexity --- after all,
much research has been dedicated to this problem;
if there were a simple solution, somebody would find it --- and perhaps
\textbf{mpdecimal} would use it.
The case of six-step FFT with its
$2^k \times 2^{k+1}$ or $2^{k+1} \times 2^k$ matrices must be an important enough
application for a matrix transposition.

Well, as it turns out, there \textit{is} a simple solution,
even more general than we need: it works for
$n \times Cn$ or $Cn \times n$ matrices, for any constant $C.$
It requires $Cn = O(n) = O(N_1) = O(N_2)$ additional memory,
has $O(N_1 N_2)$ time complexity, and is very cache-friendly,
assuming a cache-friendly procedure for transposing a square matrix is available.

\subsection{Algorithm for matrix transposition}
\label{section:mat_trans_algo}

We will now describe our algorithm for in-place transposition of $n \times Cn$ matrix,
first for $C=2.$
We will then show how to perform in-place transposition of $2n \times n$ matrix,
which is the inverse operation.
Finally, we discuss how our approach can be generalized for arbitrary $C.$

Suppose we want to transpose a matrix with $n$ rows and $2n$ columns in-place.
Let $M$ be the matrix we want to transpose.

We split $M$ into two $n$-by-$n$ submatrices, and name the left submatrix $A$ and the right
submatrix $B$:
$$
M = \begin{bmatrix}
A & B
\end{bmatrix}.
$$

Note that
$$
M^T =
\begin{bmatrix}
    A & B
\end{bmatrix}^T =
\begin{bmatrix}
    A^T \\
    B^T
\end{bmatrix}.
$$

Define $\Psi(M)$ to be the result of transposing the left and right submatrices of $M$:
$$
\Psi(M) = \begin{bmatrix}
    A^T & B^T
\end{bmatrix}.
$$
Note $\Psi(M) \ne M^T;$
we now want to explore how exactly $\Psi(M)$ differs from $M^T$ when both are written as ``flat''
sequences in row-major order.

Let us refer to the rows of $A^T$ as $\alpha_1, \ldots, \alpha_n;$
and to the rows of $B^T$ as $\beta_1, \ldots, \beta_n.$
We write $\longleftrightarrow$ for ``equivalent in row-major order flat form to''.
Then,
$$
\Psi(M) = \begin{bmatrix}
    A^T & B^T
\end{bmatrix} =
\begin{bmatrix}
    \alpha_1 & \beta_1 \\
    \ldots & \ldots \\
    \alpha_n & \beta_n
\end{bmatrix}
\longleftrightarrow
\langle \alpha_1, \beta_1, \alpha_2, \beta_2, \ldots, \alpha_n, \beta_n \rangle.
$$
It differs from
$$
M^T =
\begin{bmatrix}
    A^T \\
    B^T
\end{bmatrix} =
\begin{bmatrix}
    \alpha_1  \\
    \ldots \\
    \alpha_n \\
    \beta_1 \\
    \ldots \\
    \beta_n
\end{bmatrix}
\longleftrightarrow
\langle \alpha_1, \alpha_2, \ldots, \alpha_n, \beta_1, \beta_2, \ldots, \beta_n \rangle.
$$

So we can propose the following algorithm for transposing $M,$ although not yet in-place:
\begin{enumerate}
    \item Interpreting the whole array as $n \times 2n$ matrix, transpose the left and right submatrices.
    \item Interpreting the whole array as $2n \times n$ matrix, permute its \textit{rows} in a certain way.
        Namely, we need to apply the permutation $\rho$ that sends a row with zero-based index
        $i, 0 \le i < 2n,$ to the zero-based index $$n \cdot (i \mymod 2) + (i \mydiv 2).$$
\end{enumerate}

But permuting \textit{rows} of a matrix means permuting \textit{columns} of the transposed matrix
in the same way. To write it in a formal way, we need to introduce some new notation.

If $\pi$ is a permutation on length-$N$ sequences,
then $\pi' X,$ where $X$ is a matrix with $N$ rows, means $X$ with rows permuted by $\pi;$
and $\pi'' Y,$ where $Y$ is a matrix with $N$ columns, means $Y$ with columns permuted by $\pi.$

We can then write formally, for any matrix $X$ with $N$ columns and permutation $\pi$ on length-$N$
sequences,
\begin{equation} \label{eq:matrix_permutation_rule}
(\pi'' X)^T = \pi'(X^T).
\end{equation}

We have:
$$\rho' \Psi(M) = M^T.$$
Substitute with $M = \rho'' L$:
$$\rho' \Psi(\rho'' L) = (\rho'' L)^T.$$
Apply~\eqref{eq:matrix_permutation_rule} to the right-hand side:
$$\rho' \Psi(\rho'' L) = \rho'(L^T).$$
Simplify, since $\rho'$ is a bijection:
$$\Psi(\rho'' L) = L^T.$$

It means that we can apply $\rho$ to \textit{columns} (as opposed to rows) \textit{before} the
transpositions of submatrices (as opposed to after) --- and still get the transposed matrix as the
result.

Application of $\rho$ to a column can be done in a cache-efficient way, and requires $2n$
additional memory.

Since $(M^T)^T = M,$ the transposition of $2n \times n$ matrix is the inverse operation on the array of
the same length; it can be done as doing the inverse operations in reverse order: apply $\rho^{-1}$
to columns, then transpose the submatrices. Although $\rho^{-1} \ne \rho,$
application of $\rho^{-1}$ to a column can also be done in a
cache-efficient way, and also requires $2n$ additional memory.

We provide the pseudocode for the case of $C=2.$
Note it uses the routine $\mathrm{TransposeSquareSubMatrix}$.
It does not have to be implemented exactly in a way we demonstrated above;
it just needs to have the same semantics.
It can be implemented in a cache-friendly way by dividing the square submatrix into
tiles, perhaps recursively, transposing the elements inside tiles, and then swapping the tiles
themselves; see~\cite{cache_square_mat_trans} for details.

\begin{procedure}[!hbtp]
\caption{PermuteRho($p$, $\zeta$, $n$)}
\DontPrintSemicolon
\KwData{
    Pointer $p$ to array of size $2n$.
    Pointer $\zeta$ to scratch buffer of size $2n.$
    Integer $n.$
}
\KwResult{
    The permutation $\rho$ is applied to the array.
    The contents of the scratch buffer is undefined.
}
\Begin{
    \For{$i = 0$ \KwTo $2n-1$}{
        $\zeta[i] \longleftarrow p[i]$\;
    }
    \For{$i = 0$ \KwTo $n-1$}{
        $p[i] \longleftarrow \zeta[2i]$\;
        $p[i+n] \longleftarrow \zeta[2i+1]$\;
    }
}
\end{procedure}

\begin{procedure}[!hbtp]
\caption{UnPermuteRho($p$, $\zeta$, $n$)}
\DontPrintSemicolon
\KwData{
    Pointer $p$ to array of size $2n$.
    Pointer $\zeta$ to scratch buffer of size $2n.$
    Integer $n.$
}
\KwResult{
    The permutation $\rho^{-1}$ is applied to the array.
    The contents of the scratch buffer is undefined.
}
\Begin{
    \For{$i = 0$ \KwTo $2n-1$}{
        $\zeta[i] \longleftarrow p[i]$\;
    }
    \For{$i = 0$ \KwTo $n-1$}{
        $p[2i] \longleftarrow \zeta[i]$\;
        $p[2i+1] \longleftarrow \zeta[i+n]$\;
    }
}
\end{procedure}

\begin{procedure}[!hbtp]
\caption{TransposeMatrixNx2N($p$, $\zeta$, $n$)}
\DontPrintSemicolon
\KwData{
    Pointer $p$ to a matrix with $n$ rows and $2n$ columns stored in row-major order.
    Pointer $\zeta$ to scratch buffer of size $2n.$
    Integer $n.$
}
\KwResult{
    The matrix is transposed.
    The contents of the scratch buffer is undefined.
}
\Begin{
    \For{$i = 0$ \KwTo $n-1$}{
        $\mathrm{PermuteRho}(p + i \cdot 2n, \zeta, n)$\;
    }
    $\mathrm{TransposeSquareSubMatrix}(p, n, 2n)$\;
    $\mathrm{TransposeSquareSubMatrix}(p + n, n, 2n)$\;
}
\end{procedure}

\begin{procedure}[!hbtp]
\caption{TransposeMatrix2NxN($p$, $\zeta$, $n$)}
\DontPrintSemicolon
\KwData{
    Pointer $p$ to a matrix with $2n$ rows and $n$ columns stored in row-major order.
    Pointer $\zeta$ to scratch buffer of size $2n.$
    Integer $n.$
}
\KwResult{
    The matrix is transposed.
    The contents of the scratch buffer is undefined.
}
\Begin{
    $\mathrm{TransposeSquareSubMatrix}(p, n, 2n)$\;
    $\mathrm{TransposeSquareSubMatrix}(p + n, n, 2n)$\;
    \For{$i = 0$ \KwTo $n-1$}{
        $\mathrm{UnPermuteRho}(p + i \cdot 2n, \zeta, n)$\;
    }
}
\end{procedure}

This approach generalizes trivially to the case of $n \times C n$ and $C n \times n$ matrices;
it would require $Cn$ additional memory.

\subsection{Algorithm for bit-reversal permutation}

See~\cite{bit_rev_perm} for overview of algorithms for bit-reversal permutation.
We use the ``XOR'' approach presented therein, which is simple and has competitive performance
for small-to-medium sequence sizes.

\subsection{Optimizing the four-step algorithm for $N=3 \cdot 2^k$}

With $N=3 \cdot 2^k,$ we interpret the sequence as a matrix with 3 rows and $2^k$ columns.

Observe that steps 1 and 2 of the direct transform:
\begin{itemize}
\item length-3 DFT on each column,
\item multiplication by $(\xi_N)^{ij},$
\end{itemize}
as well as steps 3 and 4 of the inverse transform:
\begin{itemize}
\item multiplication by $(\xi_N)^{-ij},$
\item length-3 IDFT* on each column,
\end{itemize}
can be merged --- performed in a single pass over the columns.
In fact, they \textit{should} be merged, as it produces less loads/stores and leads to better
cache utilization.

If, additionally, we need to multiply each matrix element by some factor $\Phi,$
then we can meld this operation into the column operations as well; and, as it turns out,
this can save us a few multiplications per column.
Note there is no difference at which step to multiply the matrix by a scalar $\Phi,$
because all the column operations involved are linear.

Note that we are going to need to multiply everything by $N^{-1}$ after performing the inverse
transform (or by another factor if the linearity trick is used; see section~\ref{section:trick}).
For the direct transform, we only need to ``multiply'' (perform $\redc$ with $R^2$)
if the linearity trick is not used.

First, consider the case of direct transform.
Suppose we need to perform length-3 DFT on each column, then multiply each matrix element indexed
$\langle i,j \rangle$
by $(\xi_N)^{ij};$ and also multiply everything by a factor $\Phi.$

Define $\Gamma = \xi_N,$ $\Lambda = (\xi_N)^{N/3}.$
This operation on a single column with index $j,$
$\langle x, y, z \rangle \mapsto \langle x', y', z' \rangle,$ can be expressed as:
\begin{subequations}
\begin{align}
    x' &= \Phi (x + y + z); \\
    y' &= \widetilde{\Phi}_1 (x + y \Lambda   + z \Lambda^2); \\
    z' &= \widetilde{\Phi}_2 (x + y \Lambda^2 + z \Lambda),
\end{align}
\end{subequations}
where
\begin{subequations}
\begin{align}
    \widetilde{\Phi}_1 &= \Phi \Gamma^j; \\
    \widetilde{\Phi}_2 &= \Phi \Gamma^{2j}.
\end{align}
\end{subequations}

Note that $\widetilde{\Phi}_1$ and $\widetilde{\Phi}_2$ can be computed incrementally,
with two variables initially having the value of $\Phi$ and get multiplied by
$\Gamma$ and $\Gamma^2,$ correspondingly, after each iteration of the loop as we operate on columns
$0, \ldots, 2^k-1.$ This is an improvement over a separate multiply-by-$\Phi$ step:
we have 9 multiplications per column instead of 11.
Note that if we do not need to multiply by $\Phi,$ we have 8 multiplications per column.

Consider now the case of inverse transform.
We need to multiply each matrix element indexed $\langle i,j \rangle$
by $(\xi_N)^{-ij},$ then perform length-3 IDFT* on each column; and also multiply everything by a
factor $\Phi.$

Define $\Gamma = (\xi_N)^{-1},$ $\Lambda = (\xi_N)^{-N/3}.$
This operation on a single column with index $j,$
$\langle x, y, z \rangle \mapsto \langle x', y', z' \rangle,$ can be expressed as:
\begin{subequations}
\begin{align}
    x' &= \widetilde{x} + \widetilde{y}           + \widetilde{z}; \\
    y' &= \widetilde{x} + \widetilde{y} \Lambda   + \widetilde{z} \Lambda^2; \\
    z' &= \widetilde{x} + \widetilde{y} \Lambda^2 + \widetilde{z} \Lambda,
\end{align}
\end{subequations}
where
\begin{subequations}
\begin{align}
    \widetilde{x} &= x \cdot \Phi; \\
    \widetilde{y} &= y \cdot \Phi \Gamma^j; \\
    \widetilde{z} &= z \cdot \Phi \Gamma^{2j}.
\end{align}
\end{subequations}

Again, the factors for $\widetilde{y}$ and $\widetilde{z},$
that is $\Phi \Gamma^j$ and $\Phi \Gamma^{2j},$ can be computed incrementally.
Similarly to the case of direct transform, this is an improvement over a
separate multiply-by-$\Phi$ step: 9 multiplications per column instead of 11.

\subsection{Optimizing the six-step algorithm for $N=2^k$}

Remember that the reason we might turn to using the six-step algorithm is its cache
friendliness: if we interpret the sequence as $N_1 \times N_2$ matrix, then, aside from matrix
transpositions, it only accesses memory in strides of $N_1$ and $N_2.$

With $N = 2^k,$
we need to factorize $N$ into $N_1 N_2 = N$ so that $N_1$ and $N_2$ are as close to each other
as possible.

This can be done by choosing $N_1 = 2^{\lfloor k / 2 \rfloor}, N_2 = 2^{\lceil k / 2 \rceil}.$
Note that in case of even $k,$ the matrix is square; in case of odd $k,$
$N_2 = 2 \cdot N_1$ and the matrix can be transposed with the algorithm we described in
section~\ref{section:mat_trans_algo}.

Just like with the four-step algorithm, it is desirable to merge together the steps of
multiplication by $(\xi_N)^{\pm ij}$ and multiplication by a scalar $\Phi,$
and compute the factor $\Phi (\xi_N)^{\pm ij}$ for the current element at
$\langle i,j \rangle$ incrementally.
Compared to a separate multiply-by-$\Phi$ step, this approach
lowers the number of multiplications from $3N + o(N)$ to $2N + o(N).$

\section{Low-level details}

\subsection{Modular addition, subtraction, and adjustment}

For the purposes of this subsection, ``adjustment'' means reducing a value in range $[0; 2p-1]$
modulo $p$ to be in range $[0; p-1].$

Using actual conditional jumps on assembly level for modular addition, subtraction, or adjustment,
is slow on most modern architectures:
modern processors employ branch prediction, and branches of this
sort are anything but predictable. This results in lots of costly branch mispredictions.

We describe two possible ways to resolve this problem.

\subsubsection{Bit wizardry}
\label{section:bitwiz}

We assume here that two's complement is used,
all values are $w$-bit,
and a modulo $p < 2^w / 2$ is fixed.

For a signed $w$-bit value $x,$ we define $\underline{\mathrm{ExtractSign}}(x)$ as follows:
$$
\underline{\mathrm{ExtractSign}}(x) =
\begin{cases}
0, & \text{if } x \ge 0; \\
-1, & \text{if } x < 0.
\end{cases}
$$
We can compute $\underline{\mathrm{ExtractSign}}(x)$ without branches as follows:
perform signed right shift of $x$ by $(w-1)$ bits.
This leads to filling all $w$ bits with the sign bit of $x;$
all zeros result in $0,$ while all ones result in $-1.$
We can then use this value as a mask.

For $x \in [-p; p-1],$ we define
$\underline{\mathrm{AdjustSigned}}(x)$ as follows:
$$
\underline{\mathrm{AdjustSigned}}(x) =
\begin{cases}
x, & \text{if } x \ge 0; \\
x+p, & \text{if } x < 0.
\end{cases}
$$

It can be computed without branches as follows:
$$\underline{\mathrm{AdjustSigned}}(x) = x + (p \mathbin{\&} \underline{\mathrm{ExtractSign}}(x)),$$
where $\mathbin{\&}$ denotes bitwise ``AND''.
Note that $p \mathbin{\&} 0 = 0,$ $p \mathbin{\&} -1 = p.$

The modular addition of $a,b \in [0; p-1]$ can then be expressed as
$$\underline{\mathrm{AdjustSigned}}(a+b-p).$$
The normal, ``unsigned'' adjustment of $a \in [0; 2p-1]$ is expressed as
$$\underline{\mathrm{AdjustSigned}}(a-p).$$
The modular subtraction of $a,b \in [0; p-1]$ is expressed as
$$\underline{\mathrm{AdjustSigned}}(a-b).$$

\subsubsection{Conditional moves}
\label{section:cmovs}

The x86-64 platform provides ``conditional moves'' --- instructions that
store a value in a register conditionally, depending on a flag.
They do not alter control flow and are not subject to branch prediction.
Various versions of ARM have similar capabilities, although differently named.

Since version 3.8 (latest 11.0), the Clang compiler has built-in called
\verb~__builtin_unpredictable(x)~. According to the documentation,
it ``is used to indicate that a branch condition is unpredictable by hardware mechanisms such as
branch prediction logic''. We used it to implement modular addition, subtraction, and adjustment,
which resulted in 8---13\% speedup of the overall multiplication procedure.

Curiously, \verb~__builtin_unpredictable~ does not seem to produce any effect at all:
replacing \verb~__builtin_unpredictable(x)~ with \verb~(x)~ does not affect the generated code
at all. This means that Clang is smart enough to replace the branch with a conditional move
even without our hint. But this does not mean, however, that we should just write the code with
the branch and rely on the compiler to figure it out; compiling the code with branches with GNU
GCC 10.2.0 resulted in code 40---60\% slower than the one compiled with Clang, whilst with
the ``bit wizardry'' approach, GNU GCC outputs code with performance comparable to Clang.
This is because GNU GCC does not rewrite the branch into a conditional move, assuming, instead,
that it can be predicted by the hardware.

The above should mean, if anything, that compiler optimizations and heuristics are very flacky,
and should never be relied upon.

\subsection{Answer recovery}

We want to obtain an upper bound, in machine words, for the value of $\sigma_i + c_i$
in~\eqref{eq:sigmai_and_zi}.
Define $L = (B-1)^2 M.$ By~\eqref{eq:product_inequality} and~\eqref{eq:pi_bound}, we have
$$L < p_1 p_2 < (\mu / 2)^2 = \mu^2 / 4.$$

By~\eqref{eq:ci_bound}, we have $c_i \le L;$ and by~\eqref{eq:sigmai_bound}, we have
$\sigma_i < (B-1)M \le L.$ But then
\begin{equation} \label{eq:two_word_bound}
\sigma_i + c_i < 2L < \mu^2 / 2.
\end{equation}
We see that it fits into two machine words.

Note this is far from being a tight bound in numerical sense:
in practice, the value of
$\frac{L}{(B-1)M}=B-1$ is going to be much larger than 1, meaning that the exact upper bound on
$\sigma_i + c_i$ is going to be closer to $\mu^2 / 4$ than to $\mu^2 / 2.$
But this is still a tight bound in the sense that the maximum possible $\sigma_i + c_i$ would
not fit into a single machine word.
This bound also has the advantage that it works for any $B > 1.$

\subsubsection{Division by constant}

The iterative process~\eqref{eq:sigmai_and_zi} requires calculating both quotient and remainder
of integer division.
By~\eqref{eq:two_word_bound}, the dividend generally occupies two machine words.
The divisor is $B,$ which, for our choice of bases (see section~\ref{section:bases}),
always fits into one machine word.

It is possible to use the built-in two-word types, \verb~uint64_t~ on 32-bit systems and
\verb~unsigned __int128~ on 64-bit systems, for this division;
but neither the latest Clang nor GNU GCC do optimize the division of two-word types by a
constant, as they both do for one-word types. Thus, this built-in division is slow.

We use the approach described in~\cite{div_by_const} of rewriting a division by a constant
into a sequence of cheaper operations, generally a single double-width multiplication,
some bit shifts, and, for some divisors, addition and subtraction.
For each possible value of $B$ (see, again, section~\ref{section:bases}: we support 3 different
bases on 32-bit systems and 4 different bases on 64-bit systems), we generate the code for answer
recovery that performs division by $B.$

\subsubsection{Chinese remainder theorem}

The explicit solution of finding the remainder $m = ( x \mymod (p_1 \cdot p_2) ),$
where
\begin{subequations}
\begin{align}
x \mymod p_1 &= a_1; \\
x \mymod p_2 &= a_2,
\end{align}
\end{subequations}
is given by
\begin{equation}
m = a_1 + p_1 \cdot \Big( \big( r \cdot (a_2 - a_1) \big) \, \mymod \, p_2 \Big),
\end{equation}
where
$$r \equiv p_1^{-1} \, (\mymod \, p_2).$$

If we sort the primes so that $p_1 < p_2,$
then $( (a_2 - a_1) \mymod p_2 )$
is just a subtraction modulo $p_2.$
We define $\psi = (a_2 - a_1) \mymod p_2.$

Note that we can pre-calculate the value of $r.$ We can also pre-calculate
the Montgomery representation (see section~\ref{section:montgomery_intro})
of $r$ for the second modulo,
$$
\widetilde{r} = (R \cdot r) \mymod p_2,
$$
where $R = 2^w \in \mathbb{N}.$
Then, the value of $( (r \cdot \psi) \mymod p_2 )$ can be calculated as
$$\varphi = \redc(\widetilde{r}, \psi),$$
where the $\redc$ is done for the modulo $p_2.$

The value of $m = a_1 + p_1 \cdot \varphi$ is then computed as usual.

\subsection{Shenanigans with Montgomery reduction}

The idea for this optimization is due to~\cite{shenanigans}.

Remember the procedure $\mathrm{MontgomeryReduce}(a)$ that we defined in
section~\ref{section:montgomery_intro}.
It requires $$0 \le a < \mu p,$$ where $p$ is our prime modulo,
$\mu = 2^w,$ $w$ is the length of machine word in bits.
In order to compute $\redc(x,y),$ we call $\mathrm{MontgomeryReduce}(xy).$
Normally, we have $0 \le x,y < p.$
By~\eqref{eq:pi_bound}, we also have $p < \mu/2.$
It means that $$a = xy < p \mu / 2.$$

This means that we can call $\mathrm{MontgomeryReduce}(xy)$ if either $x$ or $y$ (but not both)
is in ``unadjusted'' form. Being ``unadjusted'' means being in $[0; 2p-1],$
thus possibly having different representation than the ``canonical'' one for this value
(in $[0; p-1]$).

We can obtain such an ``unadjusted'' value by omitting an adjustment:
\begin{itemize}
    \item After an addition: the unadjusted addition of $x,y \in [0; p-1]$ is just $x + y.$
    \item After a subtraction: the unadjusted subtraction of $x,y \in [0; p-1]$ is $x - y + p.$
    \item After $\redc$: observe that $\mathrm{MontgomeryReduce}$ itself
          performs adjustment as the last step.
\end{itemize}

Note that we use this optimization much more sparingly than we theoretically could, because
allowing non-local propagation of ``unadjustment'' would be a nightmare from the standpoints of
reliability and debuggability: we would need to keep track of what can, and what can not, be
``unadjusted'' at each step.

\section{Prospects}

We see the following possible ways to further improve the performance of decimal multiplication.

\begin{enumerate}
  \item \textbf{Improve cache utilization}.
    \begin{enumerate}
      \item Use breadth-first ordering, perhaps using explicit recursion.
            This can be beneficial even considering the function call overhead,
            as it leads to better cache utilization~\cite{fftw}.
      \item Explore what~\cite{gmp_paper} calls the ``Belgian approach''; see~\cite{belgian}.
      \item Explore using higher-radix transforms (raidx-4, radix-8, etc).
            Note that radix-4 is beneficial for complex FFT because,
            for a standard representation by a pair of floating-point values,
            multiplication by $i$ or $-i$ can be done significantly faster than normal
            multiplication; but finite fields of integers modulo $p$ with standard representation
            lack this property.
            Nevertheless, higher-radix transforms might be of service because they have better cache
            locality~\cite{fftw}.
    \end{enumerate}
  \item \textbf{Additional shenanigans}.
    \begin{enumerate}
      \item It is suggested in \cite{shenanigans, mont_sp} that we use primes $p$ such that $p < \mu/4.$
          Then, the Montgomery multiplication can be implemented in such a way that if the inputs
          are in $[0; 2p-1],$ then the outputs are also in $[0; 2p-1],$ and no final conditional
          subtraction is performed.
      \item It is stated in~\cite{shenanigans_presentation} that we could benefit from
          higher-radix transforms by further eliminating reductions from the butterflies.
      \item Explore signed Montgomery reduction, as defined in~\cite{signed_montgomery}.
    \end{enumerate}
  \item \textbf{Smooth the stairs}, possibly utilizing ideas from~\cite{smooth_the_stairs}.
        Currently, we can only perform transforms of lengths $2^k$ and $3 \cdot 2^k.$
        It would be nice to be able to perform transforms of lengths that are in-between.
\end{enumerate}

\section{Benchmark}

We benchmarked our implementation (see section~\ref{section:avail} for the link to the repository)
against \textbf{mpdecimal}.

Since both rely on Cooley-Tukey,
the graphs of time as a function of input size $n$ (in decimal digits)
would be staircase-like.
Define ``threshold input size'' for some fixed implementation as any value of $n$ where a
discontinuity occurs --- a new ``stair'' pops up.

We calculated the threshold input sizes of both our implementation and of the \textbf{mpdecimal}
library, within reasonable limits ($2176 \le n \le 3\cdot10^7$).
We merged all threshold values into a single list, sorted it, then formed a new list from the
averages of each two adjacent values in the sorted list.
We then appended the maximum threshold value plus one to the resulting list.
It is easy to see that this way nothing gets lost --- assuming that the staircase, aside from
discontinuities, is horizontal, we measure all possible cases.

For each value of $n,$ we performed $\lfloor \frac{8 \cdot 10^7}{n} \rfloor$
multiplications of numbers of $n$ decimal digits each.

We compiled everything using Clang 11.0.0, with option \verb!-O3!. The machine is
Xiaomi RedmiBook 14" 2019 JYU4203CN laptop with
Intel\textsuperscript{\textregistered} Core\texttrademark{} i3-8145U CPU @ 2.10GHz;
CPU scaling governors for all CPUs were set to ``performance''.

We used \textbf{mpdecimal} version 2.5.0-4 from Debian Bullseye repositories
for the ``amd64'' architecture.
For further information on \textbf{mpdecimal}, refer to
its homepage \url{http://www.bytereef.org/mpdecimal/index.html}.

\begin{figure}[H]
    \centering
    \include{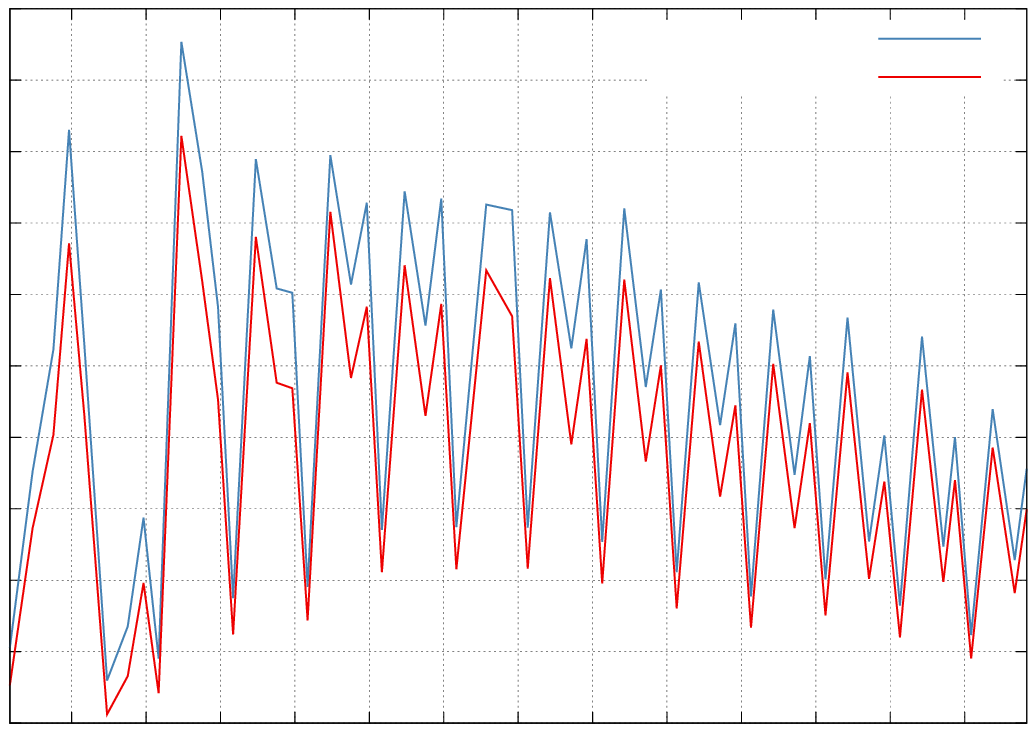}
\end{figure}

``Ratio'' is time of \textbf{mpdecimal} divided by time of our implementation.
The ``CMOVs on'' variant uses Clang-specific
\verb~__builtin_unpredictable~ built-in for modular addition, subtraction, and adjustment
(see section~\ref{section:cmovs});
the ``CMOVs off'' variant relies on ``bit wizardry'' instead (see section~\ref{section:bitwiz}).

\section{Availability}
\label{section:avail}

The code of our implementation, benchmark scripts, and \LaTeX{}
source of this paper,
are available at \url{https://github.com/shdown/decimal-multiplication-paper}.
The code is licensed under the MIT license.
The source of this paper is licensed under the Creative Commons BY 4.0 license.

\bibliography{paper}{}
\bibliographystyle{plain}

\end{document}